**Integration of AI in STEM Education – Addressing Ethical Challenges in K-12 Settings**

Authors: Shaouna Shoaib Lodhi, Shoaib Lodhi

**Abstract**

The rapid integration of Artificial Intelligence (AI) into K-12 STEM education presents transformative opportunities alongside significant ethical challenges. While AI-powered tools such as Intelligent Tutoring Systems (ITS), automated assessments, and predictive analytics enhance personalized learning and operational efficiency, they also risk perpetuating algorithmic bias, eroding student privacy, and exacerbating educational inequities. This paper examines the dual-edged impact of AI in STEM classrooms, analyzing its benefits (e.g., adaptive learning, real-time feedback) and drawbacks (e.g., surveillance risks, pedagogical limitations) through an ethical lens. We identify critical gaps in current AI education research, particularly the lack of subject-specific frameworks for responsible integration and propose a three-phased implementation roadmap paired with a tiered professional development model for educators. Our framework emphasizes equity-centered design, combining technical AI literacy with ethical reasoning to foster critical engagement among students. Key recommendations include mandatory bias audits, low-resource adaptation strategies, and policy alignment to ensure AI serves as a tool for inclusive, human-centered STEM education. By bridging theory and practice, this work advances a research-backed approach to AI integration that prioritizes pedagogical integrity, equity, and student agency in an increasingly algorithmic world.

**Keywords:** Artificial Intelligence, STEM education, algorithmic bias, ethical AI, K-12 pedagogy, equity in education

# Introduction

Artificial Intelligence has moved from science fiction to everyday reality, now invisibly powering everything from personalized Netflix recommendations to life-saving medical diagnostics (Russell & Norvig, 2016). This silent revolution is shaping how we work, learn, and connect. As AI transforms educational systems through adaptive learning technologies and automated grading systems (Touretzky et al., 2023), we are facing profound questions about the value of human judgment at the expense of algorithmic efficiencies (Binns, 2022). The classroom has become a testing ground for new technologies, revealing the tension between their potential to support personalized learning and the risk of reducing education to a collection of data points (Noble, 2018).

While AI has the potential to revolutionize personalized learning and enhance operational efficiency in classrooms (Luckin et al., 2022), its deployment raises pressing ethical concerns. A key issue is algorithmic bias – AI systems trained on historically biased datasets risk perpetuating, rather than mitigating, educational inequities (Benjamin, 2023; Noble, 2018). Predictive analytics and automated grading, if not rigorously audited for fairness, may disproportionately harm marginalized students – those historically excluded or underserved due to systemic barriers such as race, socioeconomic status, disability, language barriers, or geographic location (Adams et al., 2023; Bowers, 2017; Wang et al., 2023). Without intentional design and ongoing oversight, these technologies could reinforce systemic learning barriers instead of advancing equitable opportunities for all learners (Selwyn, 2021).

The use of AI technologies that monitor students, including facial recognition, emotion detection, and behavior tracking, can put privacy and freedom at risk (Andrejevic & Selwyn, 2020; Zuboff, 2019). Furthermore, the "black box" nature of many AI systems undermines accountability, as students, teachers, and administrators often cannot understand how decisions are made (Rahwan et al., 2022; Williamson, 2024).

To navigate these ethical complexities, it is crucial to prepare both educators and students to engage critically with AI. Teachers must be equipped to understand how AI operates, recognize potential biases, and advocate for its responsible use (Ayanwale et al., 2022; Yue et al., 2024). Simultaneously, AI literacy should be incorporated into K-12 standards, empowering students to interrogate AI's societal impacts (Long & Magerko, 2020; Touretzky et al., 2023). This goes beyond technical proficiency, fostering critical digital citizenship where students learn to question AI's ethical implications and demand transparency (Emma, 2024).

**Methodological Approach**

This paper employs a critical conceptual review and synthesis of extant literature. Our analysis is drawn from a systematic examination of recent scholarship (2016-2025) in educational technology, AI ethics, and STEM pedagogy. We prioritized peer-reviewed journal articles, seminal books, and reports from leading policy organizations (UNESCO, OECD). The objective of this methodology is not to present new empirical data but to critically analyze existing findings, identify overarching themes and tensions, and synthesize a novel, practical framework for equitable AI integration in K-12 STEM education.

Despite growing interest, the integration of AI into K-12 STEM remains underexplored, with most research focusing on generic applications rather than subject-specific implementation (Chng et al., 2023; Paek & Kim, 2021). Critical gaps persist in understanding how AI can enhance STEM learning while addressing ethical concerns like algorithmic bias (Fu et al., 2020), privacy risks in lab environments (Kamenskih, 2022), and career impacts (Subasman & Aliyyah, 2023). This paper addresses three key objectives: (1) analyzing ethical challenges in AI-driven STEM education, (2) evaluating existing AI/ethics curricula for STEM relevance, and (3) advancing a research-backed framework for responsible integration that addresses teacher readiness gaps (Casal-Otero et al., 2023) and equity concerns through STEM-specific strategies.

## Artificial Intelligence (AI) and its Educational Applications

Artificial Intelligence (AI) refers to machine-based systems capable of performing tasks that typically require human cognitive functions, including learning, reasoning, and decision-making (Russell & Norvig, 2016). Contemporary AI is primarily characterized by Narrow AI: task-specific systems that operate within constrained domains (e.g., facial recognition, recommendation algorithms) (Brynjolfsson & Mcafee, 2017). These systems are powered by *algorithms*, core computational processes that analyze data, recognize patterns, and make decisions (Yao & Zheng, 2023). Modern AI implementations predominantly utilize *machine learning (ML)*, a subset of AI that enables systems to improve performance autonomously through data analysis, often via deep learning techniques (Bengio et al., 2021; Bishop & Bishop, 2023).

In K-12 STEM education, AI-powered tools are transforming instruction. Intelligent Tutoring Systems (ITS) adapt to student responses in real-time to provide personalized feedback (Holmes et al., 2019; Miao et al., 2021), while predictive analytics use ML to identify learning gaps and shape interventions (Holstein et al., 2018). Algorithms are reshaping education through three key mechanisms: (1) personalized learning via adaptive platforms (e.g., Khan Academy) (Song et al., 2024), (2) automated assessment using Natural Language Processing (NLP) tools (e.g., automated essay scoring) (Shaik, et al., 2022), and (3) predictive analytics for early intervention (Baker & Hawn, 2022). While these implementations increase scalability, they require careful monitoring for bias and equity (Li, 2023).

**Benefits and Drawbacks of AI Applications in STEM Education**

The integration of Artificial Intelligence into STEM education is not a monolithic development but a collection of distinct technologies, each with a unique profile of pedagogical potential and ethical risk. This section analyzes the benefits and drawbacks of the three most prominent applications: Intelligent Tutoring Systems, Automated Assessments, and AI Surveillance technologies.

**Intelligent Tutoring Systems (ITS) in Education – Benefits**

Intelligent Tutoring Systems leverage AI to provide a personalized educational experience, adapting instruction in real-time to an individual student's performance. By analyzing responses, these systems dynamically adjust content difficulty and pacing, offering immediate feedback and targeted support that enables students to master challenging STEM material at their own pace (Lin et al., 2023; Luckin et al., 2022). Research demonstrates significant improvements in student achievement in complex subjects like physics (Graesser, 2016; Hu et al., 2025). Modern ITS increasingly incorporates natural language processing, virtual reality, and affective computing to create more responsive and engaging learning experiences (Wang & Jiang, 2025). Their 24/7 availability offers the potential to democratize access to quality instruction and reduce reliance on scarce human tutoring resources, thereby addressing persistent achievement gaps.

**Intelligent Tutoring Systems (ITS) in Education – Drawbacks and Limitations**

The predefined, algorithmic nature of ITS can inadvertently limit pedagogical flexibility, potentially constraining creative problem-solving and critical thinking development (Holmes et al., 2019). A primary concern is equity: these systems can perpetuate algorithmic biases embedded in their training data, disadvantaging certain student populations through biased content delivery and assessment (Baig et al., 2024; Baker & Hawn, 2022). Their standardized frameworks often fail to adequately accommodate neurodiverse learners and students with special educational needs, whose atypical learning patterns may not be recognized (Kohnke & Zaugg, 2025). Furthermore, the technology dependence of ITS exacerbates existing digital divides, creating access barriers for under-resourced schools and students (Uskov et al., 2018; Williamson, 2017). The extensive data collection required for personalization also raises significant privacy concerns, necessitating robust safeguards for sensitive student information (Regan & Jesse, 2019).

**AI-Powered Automated Assessments – Benefits**

AI-powered assessments transform evaluation in STEM education by providing adaptive, real-time analysis of student learning. Utilizing machine learning and natural language processing, these systems can analyze responses, dynamically adjust question difficulty, and deliver personalized feedback (Nazaretsky et al., 2025). They offer three key advantages: (1) delivering immediate diagnostic feedback to help students refine scientific arguments (Li, 2025), (2) automating the scoring of complex, open-ended responses (Zhai et al., 2021), and (3) identifying student misconceptions to enable targeted interventions (Luzano, 2024). Particularly beneficial for students with lower prior knowledge, AI-driven guidance has been shown to significantly improve knowledge integration and revision behaviors (Yuan et al., 2025). By combining precision with scalability, these tools enhance assessment accuracy and reduce teacher workload, freeing educators to focus on higher-order instruction (Alabdulhadi & Faisal, 2021).

**AI-Powered Automated Assessments – Drawbacks**

The primary limitation of automated assessment is its restricted capacity to evaluate nuance, creativity, or unconventional responses where human contextual interpretation is essential (Nazaretsky et al., 2025). Its reliance on historical data introduces profound risks of algorithmic bias, perpetuating existing inequalities for marginalized students and creating new inequities for those whose dialects, cultural expressions, or learning styles deviate from the AI's training norms (Davoodi, 2024; Maestrales et al., 2021). Pedagogically, over-dependence may erode teachers' formative assessment skills and encourage students to prioritize "algorithm-friendly" answers over authentic critical thinking (Lee et al., 2021). Significant implementation challenges include a lack of transparency in scoring methodologies ("black box" problem) and substantial privacy concerns regarding the continuous collection of sensitive student data (Luzano, 2024; Zhai et al., 2021). An over-reliance on standardized, immediate feedback may also potentially limit the development of higher-order cognitive skills and perseverance (Pagau & Mytra, 2023).

**AI Surveillance Technologies – Benefits**

Proponents of AI-powered monitoring tools, including facial recognition, brain-wave tracking, and behavior analysis, argue that they create responsive learning environments by enabling real-time student assessment. These technologies can analyze engagement patterns through micro-expressions, measure cognitive states like focus via EEG devices, and process digital interactions to identify learning trends (Zhang et al., 2022; Zhang, 2025; Gkintoni et al., 2025). This data is proposed to allow for dynamic content adjustment, early identification of cognitive or emotional distress, and data-driven personalization of instructional strategies, with some implementations showing improved engagement metrics.

**AI Surveillance Technologies – Drawbacks and Concerns**

Despite these purported benefits, AI surveillance raises critical ethical dilemmas that often outweigh its pedagogical potential. Privacy Violations are paramount, as continuous biometric data collection fosters a normalized culture of institutional surveillance, often deployed without meaningful student consent or robust legal safeguards (Regan & Jesse, 2019; Zuboff, 2019). Bias and Misinterpretation are rampant; algorithms frequently misread cultural or neurodiverse expressions, reinforcing stereotypes and disproportionately harming marginalized students (Zhang, 2025). Furthermore, these resource-intensive technologies exacerbate equity gaps, as underfunded schools lack the infrastructure for implementation, widening the digital divide (Li et al., 2025). These systems risk reducing the rich, contextual experience of education to a set of standardized metrics, ultimately prioritizing monitoring over meaningful learning (Zhang et al., 2022).

To conclude, while AI technologies offer transformative potential for personalization and efficiency (Luckin et al., 2022), their implementation requires careful mitigation of inherent privacy risks (Regan & Jesse, 2019), algorithmic bias (Fu et al., 2020), and equity gaps (Li et al., 2025). A balanced approach, combining AI's capabilities with essential human oversight and rigorous ethical safeguards, is critical to ensure these tools enhance rather than undermine educational equity (Selwyn, 2021).

### Ethical Concerns and Risks of AI in K-12 STEM Education

Integrating AI into K-12 STEM education presents significant ethical challenges that demand careful consideration. Recent research reveals troubling patterns of algorithmic bias in STEM learning tools.

### Epistemological Analysis of Algorithmic Bias

The bias in AI-powered STEM tools is not monolithic; it stems from interconnected sources:

1. Technical Design: Bias originates in non-diverse training datasets. For example, AI biology tools incorrectly identify specimens from less-represented environments due to this lack of diversity (Pan et al., 2025), and NLP-based assessments penalize non-dominant dialects and writing styles (Davoodi, 2024).

2. Institutional Context: Systems are often deployed in environments with existing systemic inequities. Predictive analytics can mitigate these disadvantages, for instance, by directing students from under-resourced schools toward less advanced learning paths based on historical data that reflects resource gaps, rather than student potential (Eubanks, 2018; Bowers, 2017).

3. Pedagogical Interpretation: Teachers may lack the training to critically interpret AI-generated data, potentially accepting algorithmic judgments as objective truth and thus reinforcing biased outcomes (Kohnke & Zaugg, 2025; Porhonar et al., 2025).

This bias manifests in gender disparities in AI robotics kits and the disproportionate flagging of Black and Latino students as "off-task" by behavioral analysis systems (Zeng et al., 2019).

Equally concerning are the privacy violations enabled by AI adoption. The proliferation of monitoring tools, from virtual labs to coding platforms, collects sensitive biometric data (e.g., eye-tracking, facial expressions) often without transparent consent, with data frequently flowing to third-party edtech companies and government agencies (Burkell et al., 2022; Luo et al., 2024).

Pedagogically, AI tools can impact fundamental STEM skill development. Research indicates that over-reliance on AI coding assistants can weaken debugging abilities (Yilmaz &

Yilmaz, 2023) and suppress natural scientific curiosity by reducing student questioning (Mintz et al., 2023). These cognitive impacts raise critical questions about balancing technological assistance with developing essential STEM competencies.

Furthermore, the "AI STEM divide" compounds these challenges through stark inequities in access. Schools in low-income and rural districts are significantly less likely to have the technology, infrastructure, or trained educators necessary for ethical AI integration (López Costa, 2025; Muranga et al., 2023), threatening to widen existing achievement gaps.

**Developing AI Ethics Literacy in K-12 STEM Education**

Integrating AI demands a parallel focus on developing AI ethics literacy, the knowledge and skills to critically examine AI's societal impacts. Effective approaches combine age-appropriate pedagogies with hands-on ethical problem-solving. Foundational frameworks like AI4K12's Five Big Ideas include "AI and Society" as a core pillar (Touretzky et al., 2023), while MIT's DAILy Curriculum pairs machine learning labs with justice-centered case studies (Saltz et al., 2019).

Effective pedagogical strategies emphasize active, critical engagement. Project-based learning models where students design AI solutions with ethical constraints show high critical thinking gains (Williams et al., 2023). Role-playing activities, such as simulating AI ethics boards, are also effective for developing ethical reasoning (Henry et al., 2021).

Despite promising developments, significant challenges remain. STEM educators often report low confidence in teaching AI ethics (Nazaretsky et al., 2022), and resource disparities mean schools in low-income districts have less access to quality AI ethics curricula (Muranga et al., 2023). Moving forward, implementation requires embedding ethics within existing STEM curricula, providing low-tech ethics activities for under-resourced schools, and investing in sustained teacher professional development.

## A Comprehensive Framework for Responsible AI Integration in K-12 STEM

## Theoretical Underpinnings of the Framework

The proposed framework is not merely pragmatic; it is grounded in established educational and critical theories that ensure its implementation is both effective and ethically sound. It is primarily built upon two complementary theoretical foundations:

### *The TPACK Framework*

Our tiered professional development model is explicitly designed to build teachers' Technological Pedagogical Content Knowledge (TPACK) (Chiu et al., 2021; Koehler & Mishra, 2009). This goes beyond simple technical training. The PD phases ensure educators develop a deep understanding of how AI technology (TK) intersects with specific pedagogical strategies (PK) for teaching STEM content (CK). For instance, training a biology teacher to use an AI-powered simulation (TK) is not enough; the PD focuses on *how* to use it to teach cellular respiration (CK) through inquiry-based learning (PK), thereby creating the specialized TPACK required for meaningful integration.

### *Critical Algorithmic Literacy*

The framework embeds the principle of critical algorithmic literacy (Regan & Jesse, 2019) throughout, ensuring that both students and teachers are equipped to interrogate AI systems, not just use them. This theoretical lens mandates that every phase of implementation, from pilot modules to system-wide policy, includes learning objectives focused on identifying bias, demanding transparency, and understanding the societal impacts of algorithms. This transforms AI education from a technical skill into an essential component of digital citizenship and ethical reasoning.

The synthesis of TPACK and critical algorithmic literacy ensures the framework eliminates redundancy by addressing both the *effective* integration and *ethical* scrutiny. This dual foundation bridges theory and practice, ensuring the structured progression from pilots to systemic transformation is pedagogically coherent and critically engaged.

The proposed framework addresses AI integration through three interconnected components: a phased roadmap, a professional development model, and an equity-centered approach.

**1. A Three-Phased Implementation Roadmap**

This roadmap establishes clear, scalable timelines for adoption.

- Short-Term (1-2 years): Foundational Pilots. Develop and implement 10-hour modular AI units within existing STEM subjects (e.g., dataset bias analysis in math, ethics of facial recognition in science). Evaluate through mixed-methods assessment.

- Medium-Term (3-5 years): Educator Capacity Building. Implement a tiered certification system for educators: AI Fundamentals (50 hrs.), Instructional Integration (100 hrs.), and Implementation Leadership (150 hrs.) in partnership with universities.

- Long-Term (5+ years): Systemic Transformation. Align school policy with international standards (e.g., UNESCO), implement mandatory ethics and equity audits for all EdTech, and establish robust student data governance protocols.

**Table 1: 3-Phased Implementation Roadmap**

| Implementation Phase | Key Actions | Subject-Specific Examples | Assessment & Partnerships |
|---|---|---|---|
| **Short-Term (1-2 years)** Foundational Pilots | Develop 10-hour modular AI units. Implement hands-on applications. | Mathematics: Dataset bias analysis. Science: Facial recognition ethics case studies. | Mixed-methods AI literacy evaluation. Partnership with local universities for pilot studies. |
| **Medium-Term (3-5 years)** Educator Capacity Building | Tiered certification: 1. AI | Curriculum adaptation workshops. School-wide | University partnerships for accredited courses. Pre-/post-confidence |
| | Fundamentals (50 hrs.) 2. Instructional Integration (100 hrs.) 3. Implementation Leadership (150 hrs.) | deployment strategies. | surveys for teachers. |
| **Long-Term (5+ years)** Systemic Transformation | Policy alignment with international standards. Mandatory ethics components. | Equity audits of EdTech. Student data governance committees. | UNESCO/OECD compliance monitoring. State curriculum reforms. |

## 2. A Four-Phase Professional Development Framework

This model operationalizes the roadmap through progressive teacher support.

- **Phase 1: AI Literacy (6 weeks).** Blended learning on ML fundamentals and ethical scenarios, using unplugged activities and dataset audits.

- **Phase 2: Instructional Design (8 weeks).** Professional Learning Communities (PLCs) to adapt existing STEM lesson plans and evaluate AI content.

- **Phase 3: Classroom Application (Ongoing).** Supported by co-teaching with STEM experts and reflective practice journals.

- **Phase 4: Institutionalization (Continuous).** Creating school-based AI specialist roles and conducting quarterly impact reviews.

**Table 2: Professional Development Framework**

| PD Phase | Duration | Format & Content | Key Activities | Outcomes & Evidence | Support Systems |
|---|---|---|---|---|---|
| **Phase 1: AI Literacy** | 6 weeks | Blended learning. Case study discussions. | Unplugged ML activities. Dataset bias audits. Privacy scenario training. | Measured improvement in identifying AI limitations (e.g., Lee & Perret, 2022). | Online modules. Live expert Q&A. |
| **Phase 2: Instructional Design** | 8 weeks | Professional Learning Communities (PLCs). STEM-facilitated workshops. | AI lesson plan adaptation. Collaborative content evaluation. | 3.2x increase in AI integration vs. traditional PD (Nazaretsk | Grade-level teams. Curriculum coaches. |

| PD Phase | Duration | Format & Content | Key Activities | Outcomes & Evidence | Support Systems |
|---|---|---|---|---|---|
| | | | | y et al., 2022). | |
| **Phase 3: Classroom Application** | Ongoing | In-class implementation. Reflective practice. | Co-teaching with STEM undergraduates. Weekly journaling. | 85% sustained usage rate after 1 year (Holstein & Aleven, 2022). | Teaching assistants. Just-in-time tech support. |
| **Phase 4: Institutionalization** | Continuous | Systemic integration. | School AI specialist roles. Quarterly impact reviews. | 90% program retention after 3 years. | District-level policy. Funding allocations |

## 3. Equity-Centered Implementation and Addressing Challenges

This pathway actively addresses the systemic disparities identified in the ethical analysis.

- **Low-Tech Adaptation:** Providing unplugged activities and low-resource strategies for schools lacking advanced infrastructure (Muranga et al., 2023).

- **Culturally Responsive Pedagogy:** Ensuring AI ethics discussions are responsive to local values and knowledge systems, navigating the tension between global principles and contextual morality (Eguchi et al., 2021).

- **Implementation Challenges:** A successful rollout must anticipate and plan for:

- **Institutional Resistance:** Overcoming skepticism from administrators and educators through demonstrable pilot successes and aligning AI goals with existing school improvement plans.

- **Digital Literacy Gaps:** Addressing varying levels of tech proficiency among staff and students with differentiated PD and support.

- **Urban-Rural and Socioeconomic Divides:** Advocating for policy and funding to ensure equitable access to technology and training across all districts.

- **Cultural Diversity in Ethics:** Acknowledging that perceptions of privacy, fairness, and appropriate technology use can vary across communities, necessitating inclusive and dialogic approaches to policy development.

**Implementation Scenario: A Practical Example**

Consider a mid-sized urban school district launching its short-term phase. A pilot cohort of 10 science teachers participates in a 6-week AI literacy PD (Phase 1). They then develop a 4-lesson module for their 8th-grade biology unit on ecology. Using a simplified platform, students analyze a dataset of local and global animal images to train a simple classification model. They discovered the model fails to accurately identify local species due to under-representation in the training data, a hands-on lesson in technical and geographical bias. This pilot's success, measured through student engagement surveys and pre/post AI literacy assessments, builds buy-in for the broader medium-term rollout, directly addressing potential institutional resistance.

**Conclusion**

The comprehensive framework proposed in this paper provides a structured, research-backed approach to embedding artificial intelligence literacy and competencies across K-12 school systems. By combining a three-phased implementation roadmap with a four-phase professional development model, both grounded in the theoretical foundations of TPACK and critical algorithmic literacy, this framework ensures a gradual yet systematic transition from foundational AI awareness to full institutionalization.

Key strengths of this approach include its:

1. **Evidence-Based Design**, drawing on established pedagogical theories and empirical studies to ensure effectiveness.
2. **Scalable Progression**, beginning with short-term pilots before expanding to educator certification and systemic policy alignment.
3. **Interdisciplinary integration** involves embedding AI concepts within STEM subjects while emphasizing ethical considerations and real-world applications.
4. **Sustainable Support Structures**, incorporating tiered professional development, coaching, and school-based AI specialist roles to maintain long-term adoption.

To maximize success, initial implementation should focus on well-resourced pilots and create AI fellow positions to support early adoption, while also developing low-tech strategies for under-resourced settings. Future research must examine long-term effects on student outcomes, improve assessment tools for AI literacy, and address implementation challenges such as institutional resistance and the influence of cultural diversity on ethical perceptions.

Ultimately, this framework bridges the gap between theoretical AI education principles and practical classroom application, offering a clear pathway for schools to prepare students for an AI-driven future while fostering responsible and equitable technology use. Policymakers, educators, and curriculum designers can adapt this model to their specific contexts, ensuring that AI integration in K-12 STEM education is both meaningful and sustainable.

**Final Thought – Toward Human-Centered AI in STEM Education**

The ultimate success of AI integration will not be judged solely by improved test scores but by our ability to develop learners who can critically analyze AI systems, use them ethically, and preserve human control over scientific research. This human-centered approach prepares students not just to use AI but to master it, and, more importantly, to know when not to use it. The challenge is to create STEM learning environments where artificial intelligence enhances human understanding, promotes equitable opportunities, and advances science as a tool for the collective good.

**Funding**

The author received no financial support for the research, authorship, and/or publication of this article.

**References**

Adams, C., Pente, P., Lemermeyer, G., & Rockwell, G. (2023). Ethical principles for artificial intelligence in K-12 education. *Computers and Education: Artificial Intelligence, 4*, 100131.

Alabdulhadi, A., & Faisal, M. (2021). Systematic literature review of STEM self-study related ITSs. *Education and Information Technologies, 26*(2), 1549-1588.

Aleven, V., Roll, I., McLaren, B. M., & Koedinger, K. R. (2016). Help helps, but only so much: Research on help seeking with intelligent tutoring systems. *International Journal of Artificial Intelligence in Education, 26*, 205-223.

Andrejevic, M., & Selwyn, N. (2020). Facial recognition technology in schools: Critical questions and concerns. *Learning, Media and Technology, 45*(2), 115-128.

Ashrafova, I. (2025). The Language That Rules the World: What's Behind English's Global Power? *Acta Globalis Humanitatis Et Linguarum, 2*(2), 275-283.

Athey, S. (2018). The impact of machine learning on economics. In *The economics of artificial intelligence: An agenda* (pp. 507-547). University of Chicago Press.

Ayanwale, M. A., Sanusi, I. T., Adelana, O. P., Aruleba, K. D., & Oyelere, S. S. (2022). Teachers' readiness and intention to teach artificial intelligence in schools. *Computers and Education: Artificial Intelligence, 3*, 100099.

Baig, A., Cressler, J. D., & Minsky, M. (2024). The future of ai in education: Personalized

learning and intelligent tutoring systems. AlgoVista. *Journal of AI & Computer Science, 1*(2).

Baker, R. S., & Hawn, A. (2022). Algorithmic bias in education. *International journal of artificial intelligence in education*, 1-41.

Baker, R. S., & Hawn, A. (2022). Algorithmic bias in education. *International journal of artificial intelligence in education, 32*, 1052–1092.

Bengio, Y., Lecun, Y., & Hinton, G. (2021). Deep learning for AI. *Communications of the ACM, 64*(7), 58-65.

Benjamin, R. (2023). Race after technology. In *Social Theory Re-Wired* (pp. 405-415). Routledge.

Binns, R. (2022). Human Judgment in algorithmic loops: Individual justice and automated decision-making. *Regulation & governance, 16*(1), 197-211.

Bishop, C. M., & Bishop, H. (2023). *Deep learning: Foundations and concepts.* Springer Nature.

Bowers, A. J. (2017). Quantitative research methods training in education leadership and administration preparation programs as disciplined inquiry for building school improvement capacity. *Journal of Research on Leadership Education, 12*(1), 72-96.

Brynjolfsson, E., & Mcafee, A. N. (2017). The business of artificial intelligence. *Harvard business review, 7*(1), 1-2.

Bubeck, S., Chandrasekaran, V., Eldan, R., Gehrke, J., Horvitz, E., Kamar, E., & Zhang, Y. (2023). Sparks of Artificial General Intelligence. In *Early experiments.*

Burkell, J., Regan, P. M., & Steeves, V. (2022). Privacy, Consent, and Confidentiality in Social Media Research. In *The SAGE Handbook of Social Media Research Methods* (pp. 715-725).

Cao, L. (2022). Ai in finance: challenges, techniques, and opportunities. *ACM Computing Surveys (CSUR), 55*(3), 1-38.

Casal-Otero, L., Catala, A., Fernández-Morante, C., Taboada, M., Cebreiro, B., & Barro, S. (2023). AI literacy in K-12: a systematic literature review. *International Journal of STEM Education, 10*(1), 29.

Chan, C. K. (2023). A comprehensive AI policy education framework for university teaching and learning. *International journal of educational technology in higher education, 20*(1), 38.

Chiu, T. K., Meng, H., Chai, C. S., King, I., Wong, S., & Yam, Y. (2021). Creation and evaluation of a pretertiary artificial intelligence (AI) curriculum. *IEEE Transactions on Education, 65*(1), 30-39.

Chng, E., Tan, A. L., & Tan, S. C. (2023). Examining the use of emerging technologies in schools: A review of artificial intelligence and immersive technologies in STEM education. *Journal for STEM Education Research, 6*(3), 385-407.

Chng, E., Tan, A. L., & Tan, S. C. (2023). Examining the use of emerging technologies in schools: A review of artificial intelligence and immersive technologies in STEM education. *Journal for STEM Education Research, 6*(3), 385-407.

Darling-Hammond, L. (2017). Teaching for social justice: Resources, relationships, and anti-racist practice. *Multicultural Perspectives, 19*(3), 133-138.

Davoodi, A. (2024). EQUAL AI: A Framework for Enhancing Equity, Quality, Understanding and Accessibility in Liberal Arts through AI for Multilingual Learners. *Language, Technology, and Social Media, 2*(2), 178-203.

Eguchi, A., Okada, H., & Muto, Y. (2021). Contextualizing AI education for K-12 students to enhance their learning of AI literacy through culturally responsive approaches. *KI-Künstliche Intelligenz, 35*(2), 153-161.

Emma, L. (2024). The Ethical Implications of Artificial Intelligence: A Deep Dive into Bias, Fairness, and Transparency. Retrieved from Emma, L. (2024). The Ethical Implications of Artificial Intelligence: A Deep Dive into Bias, Fairness, and Transparency.

Esteva, A., Chou, K., Yeung, S., Naik, N., Madani, A., Mottaghi, A., & Socher, R. (2021). Deep learning-enabled medical computer vision. *NPJ digital medicine, 4*(1), 5.

Eubanks, V. (2018). *Automating inequality: How high-tech tools profile, police, and punish the poor.* St. Martin's Press.

Fu, R., Huang, Y., & Singh, P. V. (2020). Ai and algorithmic bias: Source, detection, mitigation and implications. *Detection, Mitigation and Implications*.

Gkintoni, E., Antonopoulou, H., Sortwell, A., & Halkiopoulos, C. (2025). Challenging Cognitive Load Theory: The Role of Educational Neuroscience and Artificial Intelligence in Redefining Learning Efficacy. *Brain Sciences, 15*(2), 203.

Gligorea, I., Cioca, M., Oancea, R., Gorski, A. T., Gorski, H., & Tudorache, P. (2023). Adaptive learning using artificial intelligence in e-learning: A literature review. *Education Sciences, 13*(12), 1216.

Goertzel, B. (2014). Artificial general intelligence: concept, state of the art, and future prospects. *Journal of Artificial General Intelligence, 5*(1), 1.

Graesser, A. C. (2016). Conversations with AutoTutor help students learn. *International Journal of Artificial Intelligence in Education, 26*, 124-132.

Grover, S. (2024). Teaching AI to K-12 learners: Lessons, issues, and guidance. *Proceedings of the 55th ACM Technical Symposium on Computer Science Education, 1*, pp. 422-428.

Henry, J., Hernalesteen, A., & Collard, A. S. (2021). Teaching artificial intelligence to K-12 through a role-playing game questioning the intelligence concept. *KI-Künstliche Intelligenz, 35*(2), 171-179.

Holmes, W., Bialik, M., & Fadel, C. (2019). *Artificial intelligence in education promises and implications for teaching and learning.* Center for Curriculum Redesign.

Holstein, K., & Aleven, V. (2022). Designing for human–AI complementarity in K-12 education.

*AI Magazine, 43*(2), 239-248.

Holstein, K., McLaren, B. M., & Aleven, V. (2018). Student learning benefits of a mixed-reality teacher awareness tool in AI-enhanced classrooms. *Artificial Intelligence in Education: 19th International Conference, AIED 2018, 20 Proceedings, Part I 19* (pp. 154-168). London, UK: Springer International Publishing.

Hong, J. C., Li, Y., Kuo, S. Y., & An, X. (2022). Supporting schools to use face recognition systems: a continuance intention perspective of elementary school parents in China. *Education and Information Technologies, 27*(9), 12645-12665.

Hu, X., Xu, S., Tong, R., & Graesser, A. (2025). Generative AI in Education: From Foundational Insights to the Socratic Playground for Learning. *arXiv*. Retrieved from arXiv preprint arXiv:2501.06682

Hummel, P., Braun, M., Tretter, M., & Dabrock, P. (2021). Data sovereignty: A review. *Big Data & Society, 8*(1), 20539517209820.

Kamenskih, A. (2022). The analysis of security and privacy risks in smart education environments. *Journal of Smart Cities and Society, 1*(1), 17-29.

Kapur, M., & Bielaczyc, K. (2012). Designing for productive failure. *Journal of the Learning Sciences, 21*(1), 45-83.

Kohnke, S., & Zaugg, T. (2025). Artificial Intelligence: An Untapped Opportunity for Equity and Access in STEM Education. *Education Sciences, 15*(1), 68.

Kulik, J. A., & Fletcher, J. D. (2016). Effectiveness of intelligent tutoring systems: a meta-analytic review. *Review of educational research, 86*(1), 42-78.

Lee, I., & Perret, B. (2022). Preparing high school teachers to integrate AI methods into STEM classrooms. *In Proceedings of the AAAI conference on artificial intelligence*, *36*, pp. 12783-12791.

Li, H. (2023). AI in education: Bridging the divide or widening the gap? Exploring equity, opportunities, and challenges in the digital age. *Advances in Education, Humanities and Social Science Research, 8*(1), 355-360.

Li, W. (2025). Applying Natural Language Processing Adaptive Dialogs to Promote Knowledge Integration During Instruction. *Education Sciences, 15*(2), 207.

Li, Y., Tolosa, L., Rivas-Echeverria, F., & Marquez, R. (2025). *Integrating AI in Education: Navigating UNESCO Global Guidelines, Emerging Trends, and Its Intersection with Sustainable Development Goals.*

Lidwell, W., Holden, K., & Butler, J. (2010). *Universal principles of design, revised and updated: 125 ways to enhance usability, influence perception, increase appeal, make better design decisions, and teach through design.* Rockport Pub.

Lin, C. C., Huang, A. Y., & Lu, O. H. (2023). Artificial intelligence in intelligent tutoring systems toward sustainable education: a systematic review. *Smart Learning Environments, 10*(1), 41-63.

Long, D., & Magerko, B. (2020). What is AI literacy? Competencies and design considerations. *Proceedings of the 2020 CHI conference on human factors in computing systems*, (pp. 1-16).

López Costa, M. (2025). Artificial Intelligence and Data Literacy in Rural Schools' Teaching Practices: Knowledge, Use, and Challenges. *Education Sciences, 15*(3), 352.

Luckin, R., George, K., & Cukurova, M. (2022). *AI for school teachers.* CRC Press.

Luo, F., Liu, R., Nasrin, F., Awoyemi, I. D., Crawford, C., & Ma, W. (2024). Engaging students of color in physiological computing with insights from eye-tracking. *Journal of Research on Technology in Education*, 1-22.

Luzano, J. (2024). An Integrative Review of AI-Powered STEM Education. *International Journal of Academic Pedagogical Research, 8*(4), 113-118.

Maestrales, S., Zhai, X., Touitou, I., Baker, Q., Schneider, B., & Krajcik, J. (2021). Using machine learning to score multi-dimensional assessments of chemistry and physics. *Journal of Science Education and Technology, 30*, 239–254. doi:https://doi.org/10.1007/s10956-020-09895-9

McKinsey, G. I. (2023). *The state of AI in 2023: Generative AI's breakout year.* Retrieved from https://www.mckinsey.com/ai-report-2023

Miao, F., Holmes, W., Huang, R., & Zhang, H. (2021). AI and education: A guidance for policymakers. *UNESCO Publishing*.

Mintz, J., Holmes, W., Liu, L., & Perez-Ortiz, M. (2023). Artificial intelligence and K-12 education: Possibilities, pedagogies and risks. *Computers in the Schools, 40*(4), 325-333.

Mishra, R. (2024). Embracing a paradigm shift: Transitioning from traditional teaching methods to ai-based nlp education. *Research and Reviews in Literature, Social Sciences, Education, Commerce and Management, 4*, 75-79.

Muranga, K., Muse, I. S., Köroğlu, E. N., & Yildirim, Y. (2023). Artificial Intelligence and Underfunded Education. *London Journal of Social Sciences, 6*, 56-68.

Mutawa, A. M., & Sruthi, S. (2025). UNESCO's AI Competency Framework: Challenges and Opportunities in Educational Settings. *Impacts of Generative AI on the Future of Research and Education*, 75-96.

Nair, M. M., Deshmukh, A., & Tyagi, A. K. (2024). Artificial intelligence for cyber security: Current trends and future challenges. *Automated Secure Computing for Next-Generation Systems*, 83-114.

Nazaretsky, T., Ariely, M., Cukurova, M., & Alexandron, G. (2022). Teachers' trust in AI-powered educational technology and a professional development program to improve it. *British journal of educational technology, 53*(4), 914-931.

Nazaretsky, T., Mejia-Domenzain, P., Swamy, V., Frej, J., & Käser, T. (2025). The critical role of trust in adopting AI-powered educational technology for learning: An instrument for measuring student perceptions. *Computers and Education: Artificial Intelligence, 8*, 100368.

Noble, S. U. (2018). Algorithms of oppression: How search engines reinforce racism. In S. U. Noble, *Algorithms of oppression.* New York University Press.

Paek, S., & Kim, N. (2021). Analysis of worldwide research trends on the impact of artificial intelligence in education. *Sustainability, 13*(14), 7941.

Pagau, D., & Mytra, P. (2023). The Effect of Technology In Mathematics Learning. *Proximal Jurnal Penelitian Matematika Dan Pendidikan Matematika, 6*(1), 287-296.

Pan, P., Guo, S., Zhang, F., & Zhou, Z. (2025). Landmark-Based Wing Morphometrics for Three Holotrichia Beetle Species (Coleoptera, Scarabaeoidea). *Biology, 14*(3), 317.

Pedro, F., Subosa, M., Rivas, A., & Valverde, P. (2019). Artificial intelligence in education : challenges and opportunities for sustainable development. *Education 2030.* UNESCO. Retrieved from https://unesdoc.unesco.org/ark:/48223/pf0000366994

Porhonar, P., Kahtan, H., Carroll, F., & Simon, T. (2025). Bridging the Gap: Engaging Girls in Computing Through Physical Technologies. *Innovative and Intelligent Digital Technologies; Towards an Increased Efficiency, 2*, 49-62.

Prem, E. (2023). From ethical AI frameworks to tools: a review of approaches. *AI and Ethics, 3*(3), 699-716.

Rahwan, I., Cebrian, M., Obradovich, N., Bongard, J., Bonnefon, J. F., Breazeal, C., & Wellman, M. (2022). Machine Behaviour (Originally Published 2019 by Springer Nature). In S. Carta (Ed.), *Machine Learning and the City: Applications in Architecture and Urban*

*Design.*

Railing, B. P., & Bryant, R. E. (2018). Implementing Malloc: Students and Systems Programming. *Proceedings of the 49th ACM Technical Symposium on Computer Science Education*, (pp. 104-109).

Regan, P. M., & Jesse, J. (2019). Ethical challenges of edtech, big data and personalized learning: Twenty-first century student sorting and tracking. *Ethics and Information Technology, 21*, 167-179.

Russell, S. J., & Norvig, P. (2016). *Artificial intelligence: a modern approach.* Pearson.

Ryan, M., Antoniou, J., Brooks, L., Jiya, T., Macnish, K., & Stahl, B. (2021). Research and practice of AI ethics: a case study approach juxtaposing academic discourse with organisational reality. *Science and Engineering Ethics, 27*, 1-29.

Saltz, J., Skirpan, M., Fiesler, C., Gorelick, M., Yeh, T., Heckman, R., & Beard, N. (2019). Integrating ethics within machine learning courses. *ACM Transactions on Computing Education (TOCE), 19*(4), 1-26.

Selwyn, N. (2021). *Education and technology: Key issues and debates.* Bloomsbury Publishing.

Settles, B., T LaFlair, G., & Hagiwara, M. (2020). Machine learning–driven language assessment. *Transactions of the Association for computational Linguistics, 8*, 247-263.

Shaik, T., Tao, X., Li, Y., Dann, C., McDonald, J., Redmond, P., & Galligan, L. (2022). A review of the trends and challenges in adopting natural language processing methods for education feedback analysis. *IEEE Access, 10*, 56720-56739.

Sipos, R., Kutschera, A., & Klose, J. (2025). Critical Making Workshops: Sparking Meta-Discussions for Critical Thinking in Vocational Education. *Critical Education, 16*(1), 49-70.

Song, C., Shin, S. Y., & Shin, K. S. (2024). Implementing the dynamic feedback-driven learning optimization framework: a machine learning approach to personalize educational pathways. *Applied Sciences, 14*(2), 916.

StatCounter. (2025). *Statcounter GlobalStats*. Retrieved from Global search engine market share: https://gs.statcounter.com

Suárez-Guerrero, C., Rivera-Vargas, P., & Raffaghelli, J. (2023). EdTech myths: towards a critical digital educational agenda. Technology. *Technology, Pedagogy and Education, 32*(5), 605-620.

Subasman, I., & Aliyyah, R. R. (2023). The impact of technological transformation on career choices in the STEM sector. *Jurnal Kajian Pendidikan dan Psikologi, 1*(2), 129-142.

Topol, E. J. (2019). High-performance medicine: the convergence of human and artificial intelligence. *Nature medicine, 25*(1), 44-56.

Touretzky, D. S., Gardner-McCune, C., Martin, F., & Seehorn, D. (2019). K-12 guidelines for artificial intelligence: what students should know. *ISTE Conference*, *53*.

Touretzky, D., Gardner-McCune, C., & Seehorn, D. (2023). Machine learning and the five big

ideas in AI. *International Journal of Artificial Intelligence in Education, 33*(2), 233-266.

Touretzky, D., Gardner-McCune, C., Martin, F., & Seehorn, D. (2023). Envisioning AI for K-12: What should every child know about AI? *AI Magazine, 44*(2), 105-117.

Unal, Z., & Unal, A. (2025). The Impact of Professional Development on K-12 Educators' AI Integration: A Mixed-Methods Study of Attitudes, Self-Efficacy, and Implementation. *Society for Information Technology & Teacher Education International Conference* (pp. 356-371). Association for the Advancement of Computing in Education (AACE).

Uskov, V. L., Bakken, J. P., Penumatsa, A., Heinemann, C., & Rachakonda, R. (2018). Smart Pedagogy for Smart Universities. In V. H. Uskov (Ed.), *Smart Education and e-Learning 2017. 75*, pp. 3-16. Cham: Springer International Publishing. doi:https://doi.org/10.1007/978-3-319-59451-4_1

Veteška, Z. S. (2024). Transformation of Teaching through Co-Teaching and Innovative Methods. *Acta Educationis Generalis, 14*(3).

Wang, Y., & Jiang, X. (2025). AI for Education: Trends and Insights. *The Innovation*, 1-5.

Wang, Z., Saxena, N., Yu, T., Karki, S., Zetty, T., Haque, I., & Zhang, W. (2023). Preventing discriminatory decision-making in evolving data streams. *Proceedings of the 2023 ACM Conference on Fairness, Accountability, and Transparency*, (pp. 149-159).

Weerts, H., Dudík, M., Edgar, R., Jalali, A., Lutz, R., & Madaio, M. (2023). Fairlearn: Assessing and improving fairness of AI systems. *Journal of Machine Learning Research, 24*(257),

1-8.

Williams, R., Ali, S., Devasia, N., DiPaola, D., Hong, J., Kaputsos, S. P., & Breazeal, C. (2023). AI+ ethics curricula for middle school youth: Lessons learned from three project-based curricula. *International Journal of Artificial Intelligence in Education, 33*(2), 325-383.

Williamson, B. (2017). *Big data in education: The digital future of learning, policy and practice.* Retrieved from https://digital.casalini.it/9781526416346

Williamson, B. (2024). The social life of AI in education. *International Journal of Artificial Intelligence in Education, 34*(1), 97-104.

Wu, Y. (2024). Revolutionizing Learning and Teaching: Crafting Personalized, Culturally Responsive Curriculum in the AI Era. *Creative Education, 15*(8), 1642-1651.

Yao, K., & Zheng, Y. (2023). Fundamentals of machine learning. In *Nanophotonics and machine learning: Concepts, fundamentals, and applications* (Vol. 241). Cham: Springer International Publishing.

Yilmaz, R., & Yilmaz, F. G. (2023). Augmented intelligence in programming learning: Examining student views on the use of ChatGPT for programming learning. *Computers in Human Behavior: Artificial Humans, 1*(2), 100005.

Yuan, C., Xiao, N., Pei, Y., Bu, Y., & Cai, Y. (2025). Enhancing Student Learning Outcomes through AI-Driven Educational Interventions: A Comprehensive Study of Classroom Behavior and Machine Learning Integration. *International Theory and Practice in*

*Humanities and Social Sciences, 2*(2), 197-215.

Yue, M., Jong, M. S., & Ng, D. T. (2024). Understanding K–12 teachers' technological pedagogical content knowledge readiness and attitudes toward artificial intelligence education. *Education and information technologies*, 1-32.

Zawacki-Richter, O., Marín, V. I., Bond, M., & Gouverneur, F. (2019). Systematic review of research on artificial intelligence applications in higher education–where are the educators? *International journal of educational technology in higher education, 16*(1), 1-27.

Zeng, Y., Lu, E., Sun, Y., & Tian, R. (2019). Responsible facial recognition and beyond. *arXiv preprint*. Retrieved from arXiv:1909.12935

Zhai, X., Krajcik, J., & Pellegrino, J. W. (2021). On the validity of machine learning-based next generation science assessments: A validity inferential network. *Journal of Science Education and Technology, 30*, 298–312. doi:https://doi.org/10.1007/s10956-020-

Zhai, X., Yin, Y., Pellegrino, J. W., Haudek, K. C., & Shi, L. (2020). Applying machine learning in science assessment: a systematic review. *Studies in Science Education, 56*(1), 111-151.

Zhang, A. (2025). Human computer interaction system for teacher-student interaction model using machine learning. *International Journal of Human–Computer Interaction, 41*(3), 1817-1828.

Zhang, X., Zhang, X., & Dolah, J. B. (2022). Intelligent classroom teaching assessment system

based on deep learning model face recognition technology. *Scientific Programming, 2022*, 1851409. doi:https://doi.org/10.1155/2022/1851409

Zuboff, S. (2019). Surveillance capitalism and the challenge of collective action. *New labor forum, 28*(1), 10-29.